\documentclass[referee]{raa}            

\usepackage{graphicx,times}             
\usepackage{natbib}
\usepackage{amssymb,amsmath}
\usepackage{chemfig}
\usepackage{enumerate}
\bibpunct{(}{)}{;}{a}{}{,}

\usepackage{lineno}
\usepackage[pagebackref=true]{hyperref}

\begin{document}

  \title{ Mock Observations for the CSST Mission: HSTDM--Synthetic Data Generation}

   \volnopage{Vol.0 (20xx) No.0, 000--000}      
   \setcounter{page}{1}          
   \author{  SiYuan Tan 
	  \inst{1,2} 
	\and WenYin Duan
	  \inst{1,2}   
	\and YiLong Zhang
	  \inst{1}
	\and YiPing Ao 
	\inst{1}	
	\and Yan Gong\inst{1}     
   \and ZhenHui Lin
	  \inst{1}
	\and Xuan Zhang
	  \inst{1,2}
	\and Yong Shi 
	\inst{3,4}
	\and Jing Tang
	\inst{5}  
	\and Jing Li 
		 \inst{1} 
	\and RuiQing Mao
	  \inst{1} 
	\and Sheng-Cai Shi 
      \inst{1*}  
   }

   \institute{Purple Mountain Observatory, Chinese Academy of Sciences, 
   Nanjing 210023, China; $^\ast$Email: {\it scshi@pmo.ac.cn}\\
        \and
		School of Astronomy and Space Science, University of Science and Technology of China, 
		Hefei 230026,China;\\
		\and
		Department of Astronomy, Westlake University, Hangzhou 310030, Zhejiang Province, China;\\
		\and 
		 School of Astronomy and Space Science, Nanjing University, Nanjing 210093,  China;
		 \\
		 \and 
		Key Laboratory of Space Astronomy and Technology, National Astronomical Observatories, 
		Chinese Academy of Sciences, Beijing 100101, China; \\
\vs\no
   {\small Received 20xx month day; accepted 20xx month day}}

\abstract{ 
The High Sensitivity Terahertz Detection Module (HSTDM), a key component of the backend modules 
on board the China Space Station Telescope (CSST), will offer great opportunities 
for the discovery of Terahertz Astronomy, with implications 
 that extend well beyond China to the global astronomical community. 
 It is imperative that the raw data collected by HSTDM undergoes meticulous calibration and processing 
 through the HSTDM data processing pipeline (HSTDM pipeline for short) to ensure the accuracy and 
 effectiveness of the final science data to be archived for further research.
 This process necessitates that the HSTDM pipeline address instrumental artifacts and effects 
 as well as the coordination of data flow of the scheduled observing sequences 
 under all observing modes of HSTDM within the CSST automated processing environment.
 As the understanding of CSST HSTDM data processing develops during the pipeline 
 development stage, it becomes essential to assess the accuracy, the robustness and
 the performance of the HSTDM pipeline under all observing modes of HSTDM so that components 
 of the HSTDM pipeline be rationally added, removed, amended or extended within the modular framework. 
 In this paper, we develop practical simulation methods to facilitate this need.
The contribution of synthetic data generation of HSTDM observation includes two parts: 
1. HSTDM instrumental effect simulation based on both real testing profiles and simulated models; 
2. Observing data flow generation based on HSTDM observing mode scenario.
The simulation methods have been implemented and shown to be practical in testing the HSTDM pipeline  
during the development stage.   
}

   \authorrunning{SiYuan Tan, WenYin Duan, YiLong Zhang, et al.}            
   \titlerunning{Mock Observations for the CSST Mission: HSTDM--Synthetic Data Generation}  

   \maketitle


%
%
\section{Introduction to HSTDM}           
\label{sect:intro}

Featuring the flagship project of Chinese space astronomy, CSST 
is expected to be the largest space telescope developed by China 
in the coming years, with outstanding characteristics such as a large field of view, 
high image quality, and wide band. 
Its detection sensitivity and spatial resolution are on par with those of the globally renowned NASA/ESA 
Hubble Space Telescope (HST), but its  field of view and data acquisition capacity will 
significantly surpass HST. 
Equipped with an array of precision detection modules, CSST is poised to be highly competitive. 
 It is expected to achieve significant breakthroughs in the realms of cosmology, galaxies and 
 active galactic nuclei, galaxies and stars, astrometry, extrasolar planets, and celestial bodies 
 within our solar system \citep{gong2019cosmology,cao2018testing,zhan2021wide}. 
 
 One of the powerful modules aboard the  CSST is HSTDM \citep{zhang8563600}, and its core 
 components are two superconducting superconductor-insulator-superconductor (SIS) mixers both within 
 a working frequency range of 0.41-0.51THz (corresponding working wavelength is 590-730$\mu m$), 
 with a spectral resolution less than 100kHz, and a system noise temperature less 
 than 300K\citep{yao2020characterization}. 
 HSTDM serves as an excellent exemplar of Terahertz technology \citep{li2025terahertz},
 is designed to be used for spectral line observation, offering both high spectral 
 resolution and exceptionally high sensitivity. 
 The technology behind spectral line observations is known as heterodyne spectroscopy.
 In heterodyne spectroscopy, the incident sky signal $\nu_{sky} $ is mixed with a local oscillator (LO) at a 
 tunable frequency $\nu_{LO}$ close to $\nu_{sky} $, and passed onto the nonlinear mixer. The mixer is 
 designed to be very sensitive to the beat frequencies  $\left\lvert \nu_{LO}- \nu_{sky} \right\rvert $, 
 which denotes $\nu_{IF}$, are called intermediate frequencies (IF) that cover a frequency band with significant 
 lower frequencies than  $\nu_{sky} $, but retains the same information as in $\nu_{sky} $.
 The IF band signals are then amplified and passed onto the spectrometer to get the final raw 
 spectrum data for further processing.

 The signals emanating from the cosmos in the 0.41-0.51 THz frequency band contain spectral 
 signatures that reveal a range of important interstellar atomic, molecular, 
 and atmospheric tracers, including \chemfig{CI}, \chemfig{H_{2}O}, \chemfig{O_{2}}, 
 \chemfig{NH_{3}}, \chemfig{CO}, \chemfig{CS}, \chemfig{SO} and others.
 There have been some space missions worldwide that carried detection modules targeted at similar 
 frequency band, such as Odin satellite (486-504 GHz, 541-580 GHz)\citep{HJALMARSON2004504},
 Submillimeter Wave Astronomy Satellite (SWAS) (lower frequency band: 487-493 GHz)\citep{MELNICK20022051}, 
  Herschel HIFI (Band 1: 480-640 GHz)\citep{roelfsema2012orbit}. 
 The HSTDM will complement Herschel HIFI in the frequency band of 410-480 GHz, 
 with scientific objectives in two aspects:
 \begin{itemize}
\item The evolution of cosmic carbon: 
 mainly to observe the 0.492 THz neutral carbon (CI) line emission from gas clouds 
 in the Milky Way and neighboring galaxies, and to understand the distribution, 
 dynamic characteristics, 
  relationship with the environment 
  and the process of atom-to-molecule transformation.
\item Molecular spectral line survey: to obtain chemical composition of 
   different kinds of celestial bodies  in the star-forming regions 
   of the Milky Way (such as Orion-KL, Sgr B2, IRAS 16293-2422, etc.), 
late-type stars (such as IRC+10216, etc.), planets and comets\citep{AoYipingsci}.
 \end{itemize}

 The conversion of the HSTDM raw telemetry data to scientifically usable products
  is performed by a series of standard processing steps that constitute the HSTDM pipeline. 
  This involves the recombination of telemetry segment data to
   form complete spectrum data, the coordination of the raw data flow of the 
   scheduled observing sequences under all observing modes of HSTDM,  
   the efficient correction and removal of instrumental artifacts in the observation data, 
   along with the generation of standard data products of different levels.     
  The HSTDM pipeline and all the associated data products
   were designed and developed by the HSTDM data processing team under the unified leadership 
   of CSST science data processing group. 
	The HSTDM pipeline is currently nearing the end of its development phase 
	and requries comprehensive simulation and testing method to 
	assess the accuracy, the robustness and 
   the performance of HSTDM pipeline 
   under all observing modes, and this serves the purpose of this paper.

   This article is organized as follows: Section \ref{sect:Obs} reviews the observation mode 
   of HSTDM during nominal observation, which is the basis of generating synthetic observation data flow of
   the scheduled observing sequences. Section \ref{sect:data} briefly introduces the concepts and
   structures of HSTDM pipeline. 
   Section \ref{sect:analysis} discusses the simulation method for 
   HSTDM data processing pipeline at length, with simulation experiment and discussion presented in 
   Section \ref{sect:simulation_discussion}, and our conclusions drawn in 
   Section \ref{sect:conclusion}.
\section{Observing modes}
\label{sect:Obs}
According to the technical report of HSTDM \citep{AoYipingsci}, HSTDM has two types of 
observing modes: target mode, in which the telescope
observes a fixed point in the sky, and On-The-Fly (OTF) mode, which
enables continous sky scanning in a larger region.
To better present these two observing modes in the following 
subsections, it is important to give some description of the specific operations that must 
be taken in both modes.
\begin{figure*}
	\begin{center}
		\begin{tabular}{c}
	\includegraphics[width=0.6\textwidth]{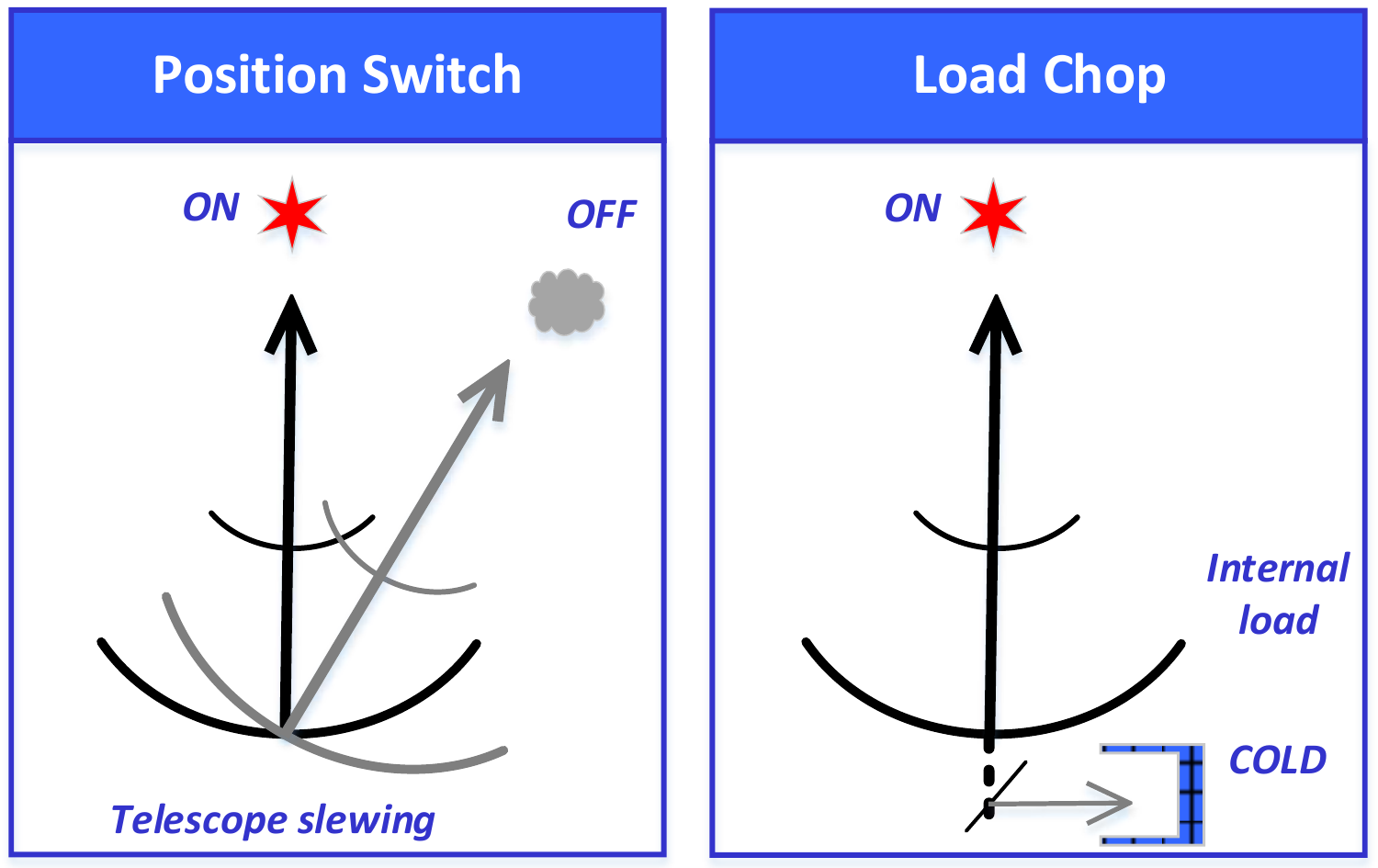}
\end{tabular}
\end{center}
	\caption{ The illustration of position switch and chop load operation during HSTDM observation.
	The position switch operation involves telescope slewing to change the actual pointing from the source 
	position (ON) to the reference position (OFF).
	The chop load  operation involves the internal mirror's adjustment to change the sky path towards 
	the internal load, which is usually a carefully engineered blackbody radiator with 
	high emissivity and known physical temperature.
	This figure is an adaptation from the HIFI observing mode illustration on page 33, 36 
	in reference \citep{Hhifi2017}.}
    \label{fig:switch_load} 
\end{figure*}

For both observing modes during nominal observation and the on-orbit calibration (OOC) stage, 
there exists two elementary operations that HSTDM must conduct: position switch and chop load.
As illustrated in Fig \ref{fig:switch_load}, position switch refers to telescope slewing
between source and reference position. ON integration can commence when either beam of HSTDM 
is directed to the source 
position. Once the beam is redirected to the OFF position, OFF integration can initiate 
accordingly. The OFF position should be selected in close proximity to the ON position, 
yet at an angular distance of no less than  $3.6'(3\sigma )$, with $\sigma \in [1.2',1.5']$ 
represents the Full Width at Half Maximum (FWHM) of the Gaussian 
beam of HSTDM at frequency range of 0.41-0.51 THz \citep{AoYipingsci}. 
The chosen OFF position should be either free of emissions within this range or exhibit 
an already known emission profile, as also outlined in \citet{HIFIOBSMODE}.
Chop load refers to the internal mirror's adjustment, which alters the sky path to 
direct the observation towards the internal load
(cold load). This operation is essential for both the HSTDM instrumental sensitivity 
measurement and the intensity calibration of the source.
Compared with telescope slewing that requires the main mirror to move, chop load is done 
through internal motor control which takes relatively less time.
\begin{figure*}[ht!]
	\begin{center}
		\begin{tabular}{c}
	\includegraphics[width=\textwidth]{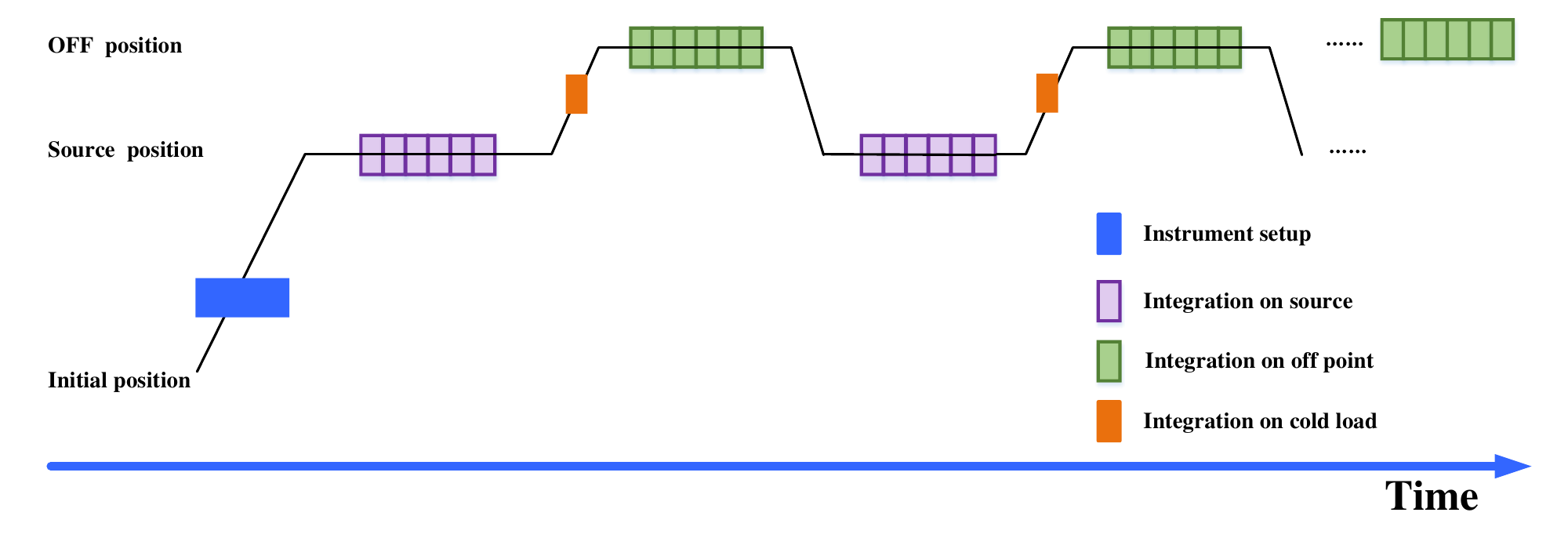}
\end{tabular}
\end{center}
	\caption{Timeline of target mode observation for HSTDM.
	Except for the initial instrument setup, the timeline consists of several cycles of basic operations such as 
	integrations at the source position, integrations on the cold load during the source to OFF 
	position slewing, and integrations at the OFF position.
	}
    \label{fig:single_m}
\end{figure*}
\subsection{Target mode}
For point source observations, HSTDM offers a dedicated target mode observation. 
This mode essentially comprises two fundamental procedural sequences:
\begin{enumerate}[1)]
\item  HSTDM  performs integration at the source position for a predefined duration, 
concurrently transferring data to the compressed storage unit of CSST. 
After an interval ascertained by the system stability time, the source integration
 is interrupted, and the system transitions to the reference integration phase via telescope slewing to the designated 
OFF position. 
\item  HSTDM assesses the instrumental sensitivity by leveraging the quantifiable disparity of the 
load measurements between the hot and cold internal loads.
	The hot and cold loads are usually  carefully engineered blackbody radiators 
	with high emissivity and known physical temperature.
	With respect to the HSTDM load measurements, we refer to those carried out with the blackbody 
	immersed in liquid nitrogen as the cold load. Conversely, the hot load corresponds to measurements taken when the 
	blackbody is placed at room temperature.
	The method to measure the instrumental sensitivity is the well known Y-Factor method\citep{tiemeijer2005improved}, which 
	can be formulated briefly as follows:
	\begin{equation}
	\begin{aligned}
		\label{equ:hot-cold}
		P_h &=(T_h+T_{sys})G \\
		P_c &=(T_c+T_{sys})G \\
		Y &=\frac{P_h}{P_c}
	\end{aligned}
\end{equation}
	\begin{align}
		\label{equ:tsys}
		T_{sys}=\frac{T_h-YT_c}{Y-1}
	\end{align}
	where $P_h$ and $P_c$ are the channel outputs of the HSTDM at hot load and cold load, respectively.
	$T_h$ and $T_c$ denote the thermodynamic temperature of the black body at hot load and cold load, 
	respectively.$Y$ represents the Y-factor,  which is determined by the measurements under both 
	hot and cold loads conditions. Additionally, $T_{sys}$ is the system noise temperature, 
	which also characterize the instrumental sensitivity of the whole system.

This load calibration should be strategically scheduled during the telescope's slew to the OFF position,  
along with other operations by leveraging observation scheduling optimization method 
like \citet{tanjatis2024}, 
thereby minimizing observational overhead as much as possible.
\end{enumerate}
These two foundational sequences are organized in a time series manner.
 A typical timeline of this mode is illustrated in Fig \ref{fig:single_m}.
  As shown in this figure, the timeline consists 
of telescope slews from the initial position to the science target,  
integrations at this position, a subsequent slew to a user-designated OFF position, and integrations at the OFF position. 
The durations of integrations at both positions is chosen to be the 
same and the sequence of pointing follows an ON-OFF patterned cycle. 
Load calibration measurements are interspersed during the telescope slews between 
the source and OFF positions, although it is not mandatory to execute load calibration after each slew.
In instances where the instrumental configuration differs from the previous setup, an instrument reset 
is required during the initial slew to the target source position, followed by a load calibration procedure.
There are five parameters that can characterize this timeline sequences, 
as listed in Table \ref{tab:single}. 
\begin{table}[h!]
	\centering
	\caption{Parameters of target mode observation timeline.}
	\label{tab:single}
	\begin{tabular}{ll} 
		\hline
		Symbols & Meanings\\
		\hline
		$\varDelta T_{d\!u\!m\!p}$ & Data dump interval (Unit: seconds); \\
		$ T_{s\!w\!i\!c\!h}$ & switch time between ON and OFF position (Unit: seconds); \\
		$N_{d\!u\!m\!p}$ & Number of data dumps per phase; \\
    $N_{O\!N-O\!F\!F}$ & Number of ON-OFF cycles;  \\
    $t_{l\!o\!a\!d}$ & Internal load period (Unit: seconds);  \\
		\hline
	\end{tabular}
\end{table}

\begin{figure*}[ht!]
	\begin{center}
		\begin{tabular}{c}
	\includegraphics[width=0.8\textwidth]{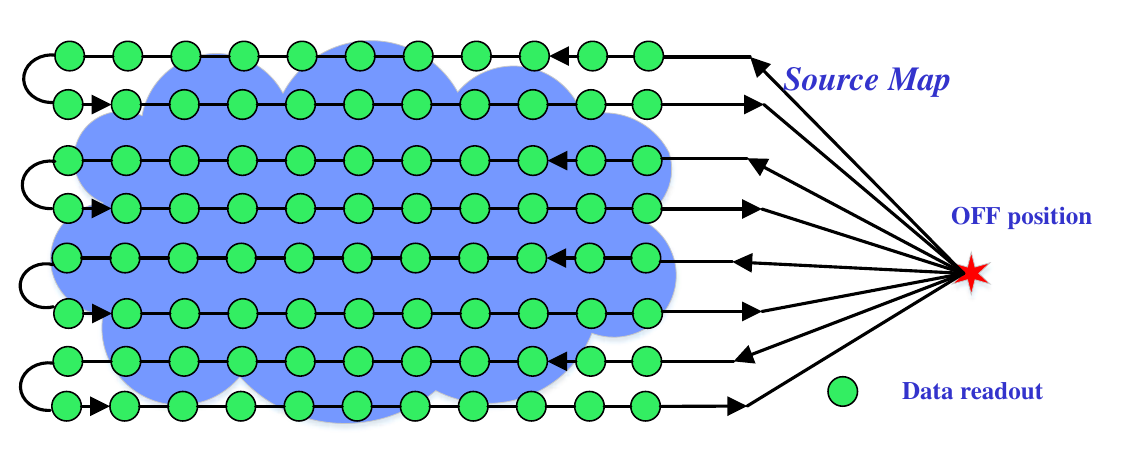}
\end{tabular}
\end{center}
	\caption{Schematic illustration of the OTF observation mode.
	The process of integrating on source map and data dumping 
	takes place as the telescope performs a scan along a particular row within the grid. 
	After completing each row, the telescope reverses its scanning direction to commence 
	the scan of the subsequent row. Integration and data dumping activities are temporarily 
	suspended during these directional transitions. Once a period has elapsed, 
	as determined by the system's stability criteria, the mapping procedure is paused 
	momentarily to conduct reference measurements at the reference (OFF) position.
	}
    \label{fig:OTF_m}
\end{figure*}
\subsection{On-The-Fly mode}\label{subsec:OTF}
\begin{figure*}[h!]
	\begin{center}
		\begin{tabular}{c}
	\includegraphics[width=\textwidth]{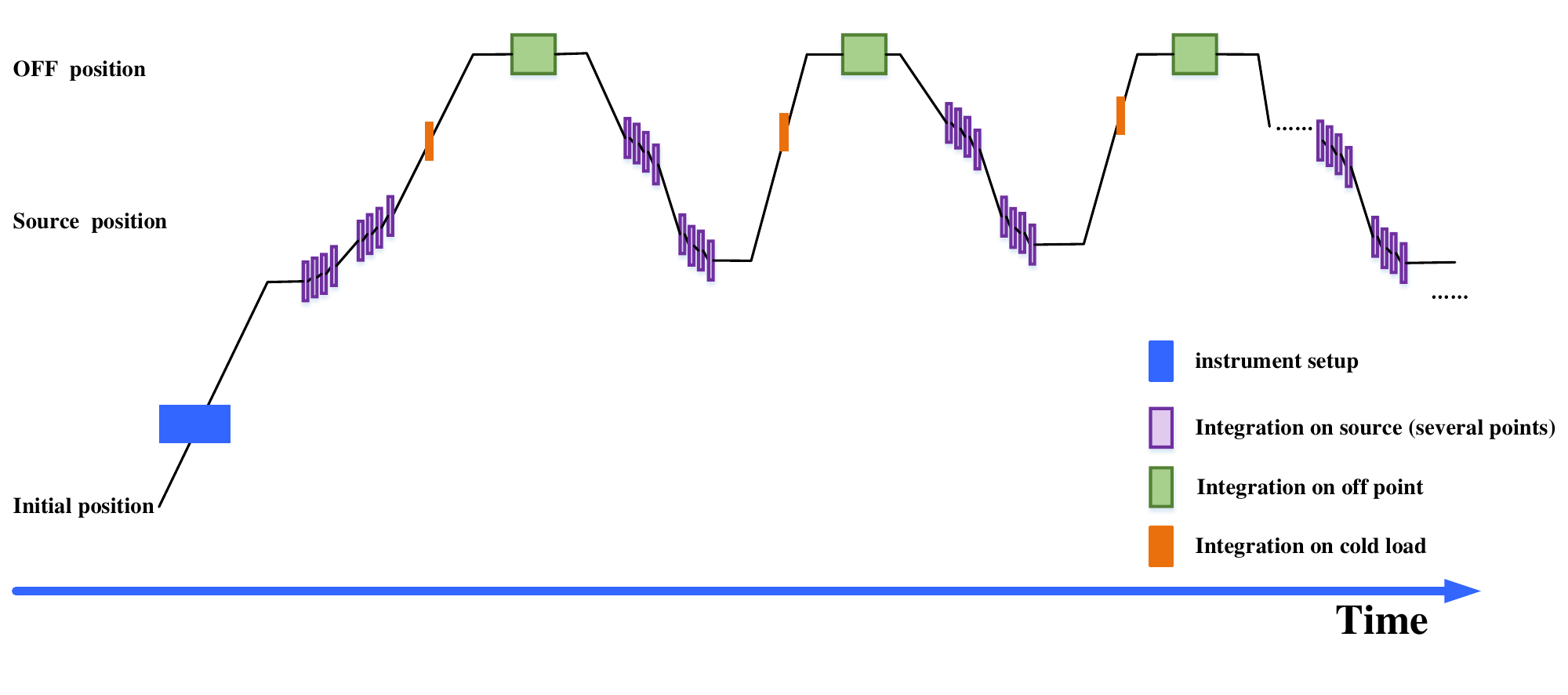}
\end{tabular}
\end{center}
	\caption{Timeline of the OTF observation for HSTDM.
	Except for the initial instrument setup, the timeline mainly 
	consists of several cycles of basic operations such as 
	integrations at source map during the scanning, a subsequent slew to a user designated OFF position, 
	integration on cold load during the slewing, and integrations at the OFF position.
	}
    \label{fig:OTF-Timeline}
\end{figure*}

For observations of extended sources, HSTDM utilizes the On-The-Fly (OTF) 
observation technique. OTF is an observing technique in which the 
telescope is driven smoothly and rapidly across a region 
of sky, or ``field'', while data and antenna position information 
are recorded continuously \citep{mangum2007fly}. The schematic view of 
typical OTF observation and its timeline sketch are presented  
in Fig \ref{fig:OTF_m} and Fig \ref{fig:OTF-Timeline} respectively.
 As depicted in Fig \ref{fig:OTF_m}, integration of the source 
occurs, and data dumps are captured during the telescope's scan of a 
specific row within the map grid. 
Following each row, the telescope reverses direction to initiate the 
scan of the subsequent row. Integration and data dumping are paused during 
these directional changes. After a duration set by the system's stability 
requirements, the mapping process is temporarily halted 
for reference measurements at the OFF position. 
Load calibration is commonly executed during the 
telescope's slew to the OFF position as illustrated in Fig \ref{fig:OTF-Timeline}.
There are totally six parameters that can characterize this 
timeline sequences, as listed in Table \ref{tab:OTF}. 
\begin{table}[hb!]
	\centering
	\caption{Parameters of OTF observation timeline.}
	\label{tab:OTF}
	\begin{tabular}{ll} 
		\hline
		Symbols & Meanings\\
		\hline
		$\varDelta T_{d\!u\!m\!p}$ & Data dump interval (Unit: seconds); \\
		$ T_{s\!w\!i\!c\!h}$ & switch time between the ON and OFF positions (Unit: seconds); \\
		$N_{O\!N}$ &  Number of ON measurements made per OFF; \\
    $N_{O\!F\!F-scan-O\!F\!F}$ & Total number of OFF-SCAN-OFF cycles;  \\
	$t_{l\!o\!a\!d}$ & Internal load period (Unit: seconds);  \\
	$V_{s\!c\!a\!n}$ & Scanning speed during ON measurements (Unit: $^{\prime \prime}/sec$);  \\
		\hline
	\end{tabular}
\end{table}

Typically, the mapped area is observed through multiple coverages, 
accumulating the total integration time per source to meet the 
necessary requirements. The RMS noise of switched measurements is given by the following:
\begin{align}
	\sigma  &=\frac{T_{sys}}{\eta_{spec}\sqrt{\Delta \nu t_{on}}}\sqrt{1+\frac{t_{on}}{t_{off}} }
\end{align}	
where $T_{sys}$ denotes the system noise temperature, $\Delta \nu$ is the spectral 
resolution of the measurement, $t_{on}$ denotes the source measurement period,
 $t_{off}$ denotes the OFF measurement period, and $\eta_{spec}$ is the 
 spectrometer efficiency. The efficiency of the OTF mode, 
 due to its high scanning velocity, significantly exceeds that of 
traditional target mode observations. The optimum duration of an OFF measurement 
for OTF is given by \citep{mangum2007fly}:
\begin{align}
	\label{equ:toff}
	t^{opt}_{off}  &=\sqrt{N_{on}}t_{on}
\end{align}	
where $N_{on}$ denotes the number of ON measurements made per OFF measurement. 
It is very clear from Equation \ref{equ:toff} that the efficiency of OTF can 
be improved with a larger $N_{on}$. However, the maximum allowable time 
between two OFF integrations is constrained by the system stability time, 
denoted as $t_{s\!t\!a\!b\!i\!l\!i\!t\!y}$. 
Meanwhile, the minimum integration time, $t_{on}$, for each source point 
is limited by the data storage rate. The scanning velocity must be 
calibrated such that the telescope's motion during a single integration 
between two data readouts covers less than half (Nyquist frequency) 
of, or even less than one third of the beam width.
This adjustment is crucial to minimize dynamic blurring in the direction of 
telescope motion while facilitating rapid map scanning. 
Furthermore, when accounting for the dead time associated with the 
telescope's slew from the source position to the OFF position, as well as 
the change in scan direction, a complex optimization process is required to 
determine these parameters effectively.    
In Fig \ref{fig:OTF-Timeline}, the scanning motion within the map is symbolized by a 
series of small, step-like structure in purple. Concurrently, at the designated OFF position, 
multiple data readouts, indicated by green rectangles, are conducted for identical locations.
 
\section{HSTDM data processing pipeline concepts}
\label{sect:data}
\begin{figure*}[ht!]
	\begin{center}
		\begin{tabular}{c}
	\includegraphics[width=\textwidth,trim=0cm 2cm 0cm 0cm]{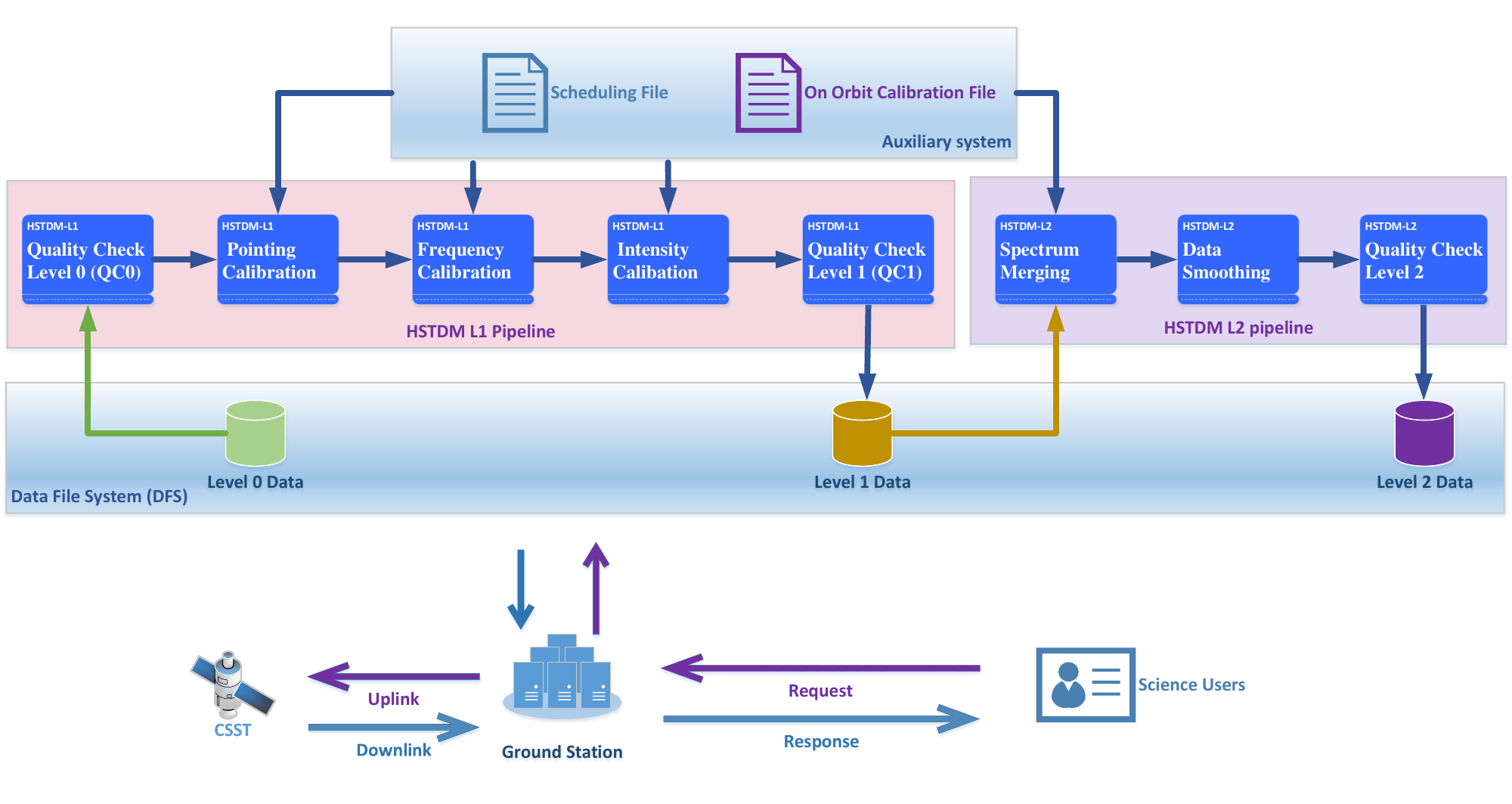}
\end{tabular}
\end{center}
	\caption{Overview of the HSTDM data processing pipeline.
	The HSTDM data processing pipeline is a two-tiered pipeline designed 
	for execution at ground stations. 
	The pipeline's inputs and outputs correspond to Level 0 data, Level 1 data products, 
	and Level 2 data products, respectively. 
	Essential for its operation are the scheduling file and the On Orbit Calibration file, 
	which are provided by the auxiliary systems.
	}
    \label{fig:hstdm-pipeline}
\end{figure*}

The CSST HSTDM does not produce high-level science data in orbit, it only generates raw 
telemetry data which is transmitted to the ground station where 
the Level 0 data is produced and archived according to the 
Interface Control Document (ICD) of HSTDM and the and HSTDM Level 0 data 
definition file \citep{csstlevel0}.
The HSTDM pipeline starts with level 0 data, and is responsible for converting 
the Level 0 data to Level 1 and Level 2 science data through two levels of 
calibration procedures that will be briefly discussed in the following 
subsections. A schematic view of HSTDM pipeline is displayed 
in Fig \ref{fig:hstdm-pipeline}. 
\subsection{Data format and Calibration}
As depicted in Fig \ref{fig:hstdm-pipeline}, calibrating HSTDM data involves 
converting a series of raw instrument outputs into scientifically usable data, 
which is typically in the form of antenna temperature versus 
frequency. Raw telemetry data from a single readout of HSTDM is 
in the form of spectral count versus corresponding channel numbers. 

An integration consists of several readouts.
Given the instrument drifts of HSTDM and the changing pointing during 
different readouts, these readouts cannot immediately co-added. 
They have to be firstly converted into level 0 data, which contains 
not only the raw readout that comes from space, but also incorporates important 
positional and state information of the CSST platform at the beginning 
and end of the specific readout, as well as the observation mode and sequence 
information at the scheduling level. In fact, the level 0 data is 
designed to be self-contained, in that necessary information for subsequent 
processing is provided.

For HSTDM, it is convenient to take observations as a function of time steps 
that are prescribed by the concrete observation command sequences based on 
two aforementioned observation modes. At any moment, HSTDM is 
executing the flow of the observation command sequence, be it 
the configuration of the HSTDM instrument, slewing the telescope, 
integrating on-source or off-source position, or doing internal load. 
At any time, HSTDM observes intensity as a function of frequency, 
and outputs spectral data with precise timestamp codes. 
A given observation sequence could have hundreds and thousands 
of integrations, their initial outputs are collected, recombined, 
restructured, and reformatted into a group of level 0 data files 
by addressing the specific protocol and scheduling information 
meticulously at the ground station. 
The generated level 0 data files are then archived in the Data File 
System (DFS) and can be accessed through the dedicated and unified interface 
provided by the CSST data processing system. The level 0 data files and 
the relevant metadata files from both the scheduling system and the on-orbit 
calibration system are all that are needed for the HSTDM pipeline. 
The HSTDM pipeline is designed in a way that it can handle data 
calibration, regardless of the specific observation mode, and outputs 
Level 1 and Level 2 data products that are in the Flexible Image 
Transport System (FITS) format with detailed definition 
in \citet{csstlevel1} and \citet{csstlevel2}respectively.  
\subsection{Processing levels}
Level 0 is the rawest form of HSTDM data available for the HSTDM pipeline. 
To begin with, Level 0 data must undergo a quality check, which consists of 
several sanity checks of FITS header items and data dimensions before 
being handed over to the HSTDM L1 pipeline. 
The main functions of Level 1 pipeline are:
\begin{enumerate}[1)]
 \item Flux-calibrate the HSTDM data using internal 
load measurements and reference measurements; 
\item Calibrate the 
 pointing of HSTDM by applying pointing model and calibration reference 
 system parameters retrieved 
 from the On-Orbit calibration system.
\end{enumerate}
 Additionally, the channel numbers of Level 0 data are converted to the 
 observed frequencies with 
 additional necessary calibration to remove the Doppler effect during the observations.
 
 The Level 2 pipeline is mainly responsible for combining and merging integration 
 data taken at different times of observations, as well as applying advanced methods 
 to mitigate the HSTDM's instrument effects including baseline 
 distortion, standing waves, and sideband imbalances, to name a few. 
 For target mode observation, the Level 2 pipeline applies the radiometric weighting 
 to each Level 1 data file in the same group, which are then co-added to obtain 
 long integration spectral data. For OTF observation, the Level 2 
 pipeline provides a standard regridding method to get the resampled data 
 of the targeted area based on a series of sampling
 point data.    
\section{Synthetic Data Generation of HSTDM observations}
\label{sect:analysis}
The current development of the HSTDM pipeline is nearing completion, except for 
certain high level calibration methods within the Level 2 pipeline that remain 
undetermined. This uncertainty stems from a lack of comprehensive data 
regarding the impact of the primary optical system of the CSST on 
the HSTDM's performance.
The Level 1 pipeline has completed two rounds of unit and integrated testing. 
Given this progress, it is now essential to conduct necessary simulation tests on 
the HSTDM pipeline before commencing the comprehensive ground testing of the CSST data 
processing pipeline. This step will follow the installation and successful hardware 
testing of all backend modules within the primary optical instrument.
These preliminary simulation tests are critical for identifying potential issues and 
optimizing HSTDM pipeline performance, thereby ensuring the accuracy, the reliability, 
and the efficiency of HSTDM pipeline in the subsequent ground testing and in-flight phases. 

\begin{figure*}[h!]
	\begin{center}
		\begin{tabular}{c}
	\includegraphics[width=\textwidth]{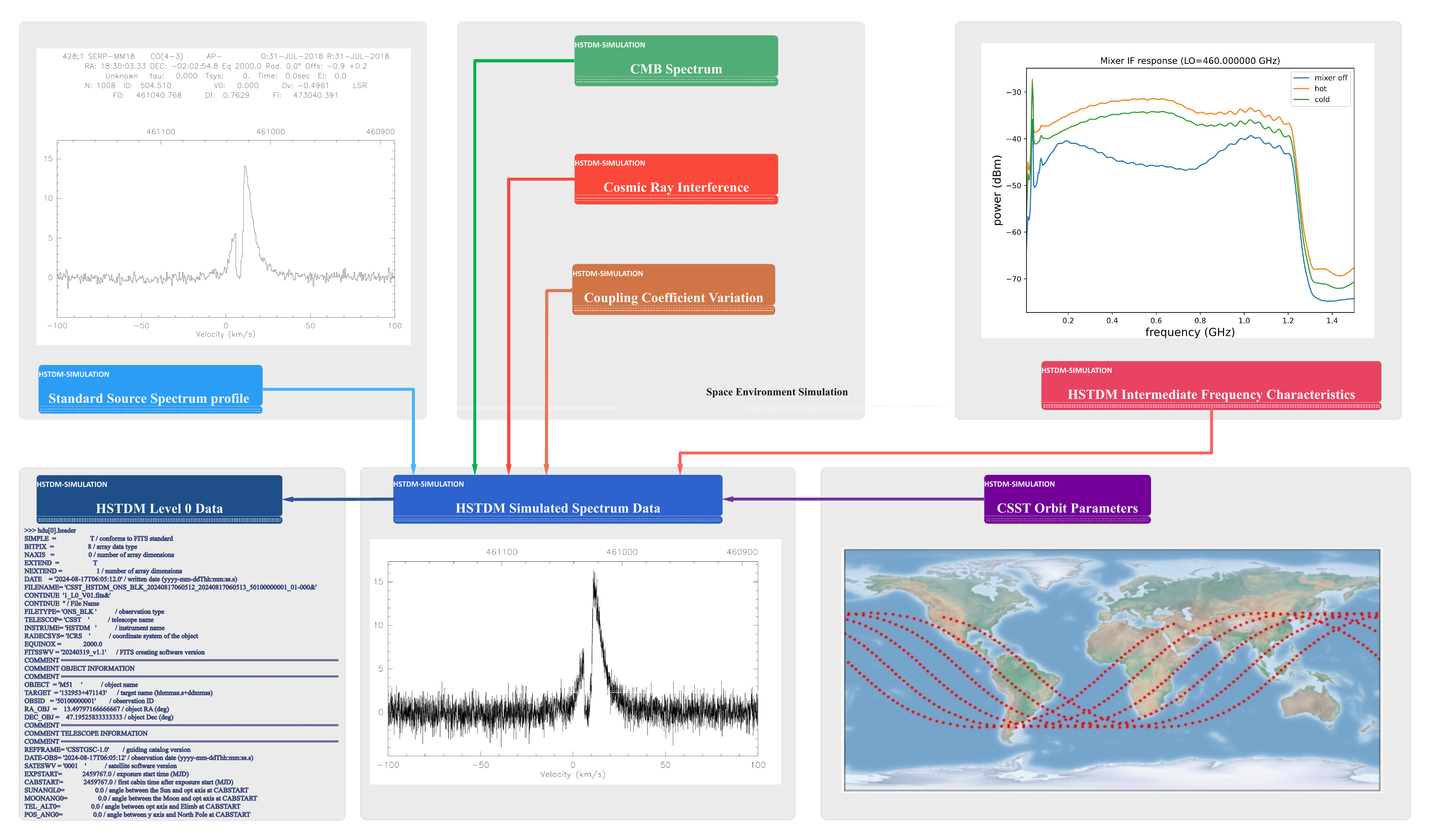}
\end{tabular}
\end{center}
	\caption{Schematic diagram of the proposed synthetic method for the HSTDM observation.
		This approach encompasses multiple stages of synthesis. Initially, a standard source 
		spectrum profile is extracted from an open archive. Subsequently, it is convolved 
		with the instrumental effects spectrum, which are estimated from the HSTDM's Intermediate 
		Frequency (IF) characteristics measurements. Following this, it is further convolved with 
		the simulated CMB spectrum and the cosmic ray interference spectrum.
		Subsequently, the synthesized data is subjected to variations in the coupling coefficient 
		based on the CSST orbit parameter. Finally, the resultant spectral data is 
		converted to the HSTDM raw output format, derived from the cold-hot measurement in 
		spectrum count units, and is encapsulated with a Level 0 data header to generate  
		the HSTDM standard level 0 data.
		The standard CO(4-3) spectrum is illustrated in the standard source spectrum profile block, 
	and SIS mixer response is illustrated in the HSTDM intermediate frequency characteristics block, 
	ground track of CSST is illustrated in the CSST orbit parameter block, and synthetic CO(4-3)
	spectrum is illustrated in the HSTDM simulated standard spectrum data block and  
	 glimpse of FITS header information of HSTDM level 0 data is displayed in the HSTDM level 0 data
	block.
	}
    \label{fig:simulate-shematic}
\end{figure*}

To facilitate this urgent need, we propose a Synthetic Data Generation method of HSTDM 
observations, focusing on simulating HSTDM instrumental effects, space environment 
conditions, and observation data flow based on observing mode scenarios. 
We also consider the intermediate frequency characteristics to obtain the basic spectrum profile, 
and the CSST orbit parameters to derive the position and velocity information at any timestamps. 
All the above mentioned aspects are necessary to generate the HSTDM's simulation data that 
resembles real-world scenarios as closely as possible.   
The schematic view of the proposed method is displayed in Fig \ref{fig:simulate-shematic}.  
We would discuss related techniques at length in the following subsection \ref{subsect:instrument_e}, 
\ref{subsect:spaced_e} and \ref{subsect:observing_m}.

\subsection{ Simulation of HSTDM Instrumental Effects}
\label{subsect:instrument_e}
HSTDM is meticulously engineered to detect incoming signals that possess exceptionally low 
power levels with ample output. This necessitates a substantial receiver gain. 
In reality however, the receiver gain is not perfectly stable. Consequently, 
even minor fluctuations in gain can contribute predominantly to the thermal noise of the 
receiver. Hence, the instability of HSTDM should be studied and considered in our simulated data.

\begin{table}[th!]
	\centering
	\caption{Summary of the configuration of collected data used for instability of HSTDM}
	\label{Tab:data_config}
	\begin{tabular}{ll} 
		\hline
		Items & Values\\
		\hline
		Temperature & low temperature ($< 8 K$); \\
		Detector &  SIS-2; \\
    Local Oscillator & On, with LO frequency 460GHz ;  \\
    $V_{SIS}$ & $8mV$ (Bias voltaga);  \\
		\hline
	\end{tabular}
\end{table} 

To characterize the instability of HSTDM, we apply Allan variance analysis \citep{riley2008handbook} of all the channels' output 
data of HSTDM that are collected during the integration testing phase of HSTDM qualification components.
The main configuration of the testing environment are summarized in Table \ref{Tab:data_config}

We use the open source project AllanTools \citep{allantools} to calculate the Overlapping 
Allen deviation (OAD)  of HSTDM output $x_{n,i}$, 
where $i$ denotes the channel number, $x_{n,i} $ is the time-series of the $i_{th}$ 
channel output, spaced by the measurement interval $\tau_0$, with length $N$. 
The OAD of $x_{n,i}$ can be formulated as:
\begin{align}
	\sigma_{oadev}^{2}(m\tau_0 )=\frac{1}{2(m\tau_0)^2(N-2m)}\sum_{n=1}^{N-2m}(x_{n+2m,i}-2x_{n+m,i}+x_{n,i})^2
\end{align}
where $\sigma_{oadev}^{2}(m\tau_0 ) $ is the OAD of $x_{n,i}$

\begin{figure*}[ht!]
	\begin{center}
		\begin{tabular}{c}
	\includegraphics[width=\textwidth]{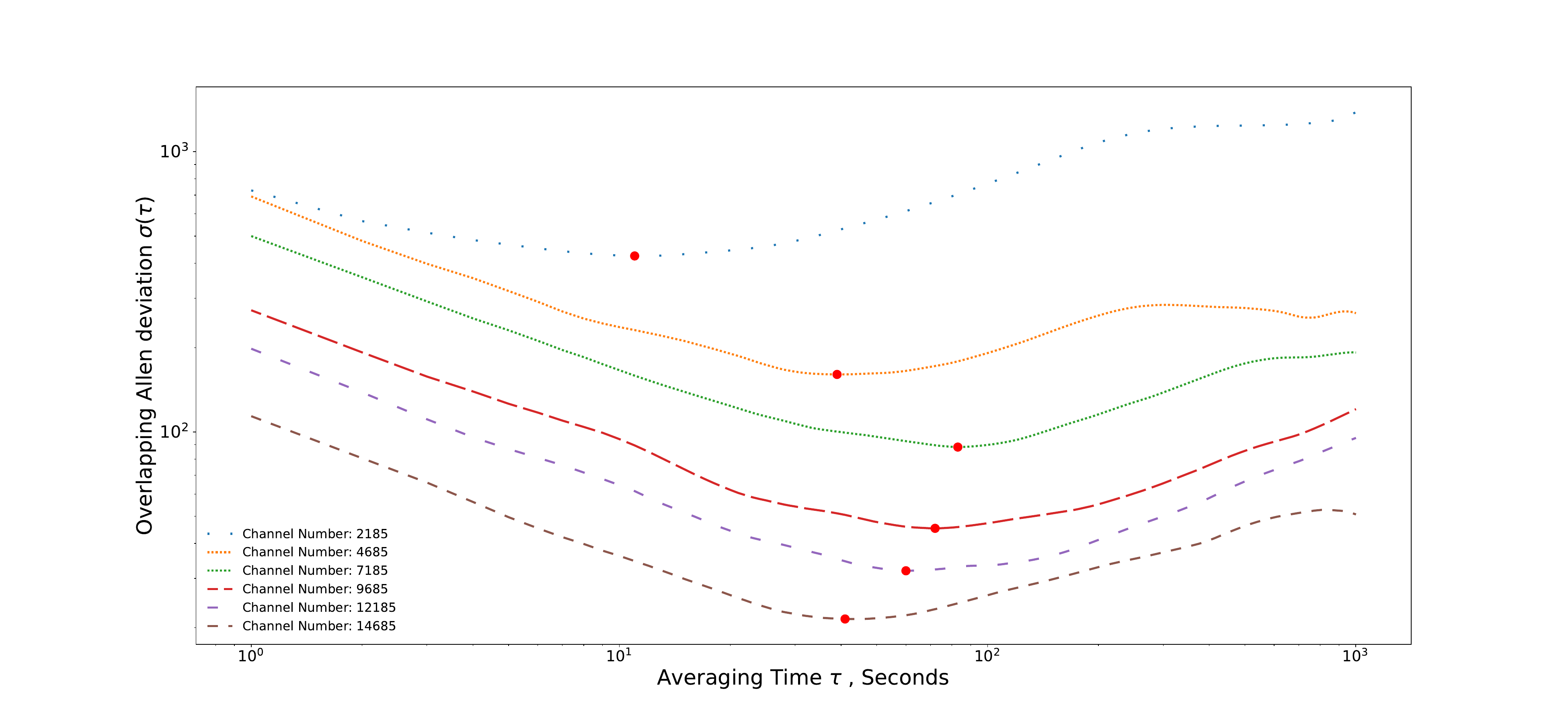}
\end{tabular}
\end{center}
	\caption{The Overlapping Allen deviation of the output from six randomly selected 
	channels, with respect to the averaging time $\tau$, spans a range from 1 second to 1000 seconds.
	 Allan time $t_A$ of each channel output is determined by the value $\tau$ that 
	 yields the minimum of its corresponding OAD $\sigma(\tau) $. The Allan time $t_A$ of    
	 the output of these six randomly selected channels is distinctly marked with red circles on each curve.
	}
    \label{fig:hstdm_allan}
\end{figure*}

The OAD of 6 randomly selected channel outputs of HSTDM under the testing environment is shown 
in Fig \ref{fig:hstdm_allan}. 

As the figure illustrates, the OAD of each channel output is much different from each other, 
although all seem to have similar trend of OAD value changes. The turning points in 
red circles in $\tau$ values, where the corresponding OAD reaches the minimal value is 
what we called the Allan time $t_A$, or system 
stability time as we have mentioned in subsection \ref{subsec:OTF}.
 We observed that $t_A$ of the 6 displayed channel outputs deviate from 10 seconds 
 to 90 seconds, indicating the difference of instability of channel output.
Totally, there are 16384 channels, which correspond to 0-1.2 GHz in the intermediate frequency 
range of each of two HSTDM detectors, with the range in 0-0.16 GHz and 1.16-1.20 GHz 
omitted due to performance deterioration. Therefore, our interested channel numbers 
are from 2185 to 15838.

\begin{figure*}[ht!]
	\begin{center}
		\begin{tabular}{c}
	\includegraphics[width=\textwidth,trim=0cm 2cm 0cm 1cm]{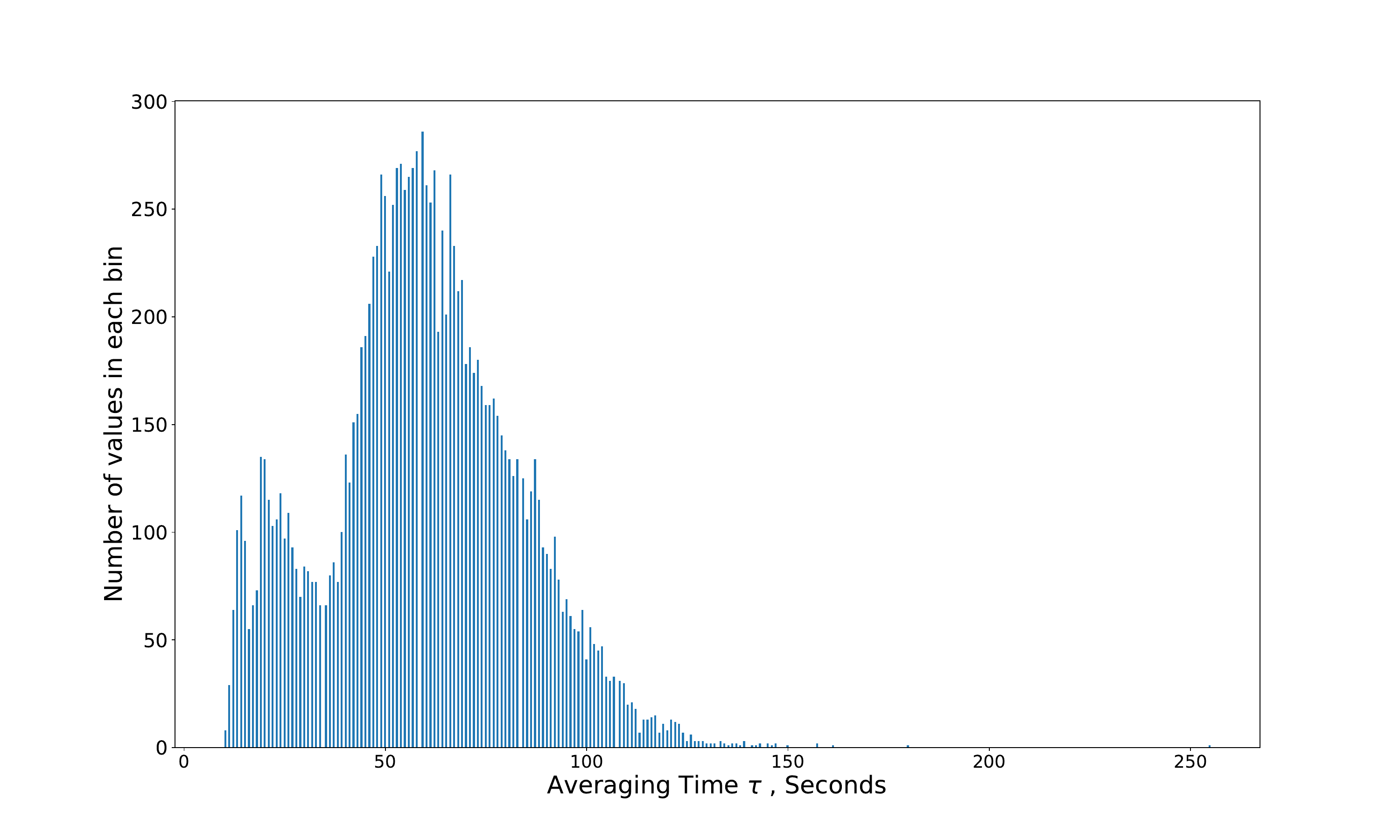}
\end{tabular}
\end{center}
	\caption{The histogram plot of $t_A$ for the channel output from channel 2185 to 15838. 
	Although the total channel number of HSTDM is 16384, the effective data 
	outputs are derived from channel 2185 to 15838. The Allen time $t_A$ for
	the majority of these 13653 channel outputs spans a range of 10 to 100 seconds.
	}
    \label{fig:allan_hist}
\end{figure*}

To further uncover the difference of $t_A$ between all our interested channel output, 
we use histogram to display the distribution of all the $t_A$ that are calculated. 
The results are shown in Fig \ref{fig:allan_hist}. 

As the figure presents, we observe that the $t_A$ values for the output from 
channel 2185 to channel 15838 is from 10 seconds to 255 seconds,
 with a median value of 59 seconds. Thus, from the statistics respective, 
 we choose $t_A =60$ seconds in our simulation.
 $t_A$, also refers to $t_{s\!t\!a\!b\!i\!l\!i\!t\!y}$, is an important parameter 
 both to characterize the stability of HSTDM and to determine 
 the concrete values of parameters concerning the observation mode profile 
 that will be discussed in subsection \ref{subsect:observing_m}. 

To simulate the instability effects in the simulated output data, we model the measurement $X_{it}$ in the output of channel $i$ 
as an Ornstein-Uhlenbeck (OU) process  \citep{caceres1997generalized}, which can be modeled by the following 
stochastic differential equation (SDE): 
\begin{align}
	dX_{it} = -\theta(X_{it} - \mu)dt+\sigma dW_t \label{equ:ou_diff}
\end{align}
where, $\mu$ is the long-term mean or the equilibrium level to which the process 
reverts, $\theta >0$ is the rate at which the process reverts to the 
mean $\mu$, $\sigma > 0$ is the volatility or the deviation of the random fluctuations.
$W_t$ is a standard Wiener process(i.e. a process for which $dW_t/dt=W'_t$ is a white 
noise process), which represents Brownian motion.
The analytical solution to the $X_{it}$ can be formulated as the following:
\begin{align}
	X_{it+\tau}=(1-e^{-\theta \tau} )\mu +X_{it} e^{-\theta \tau} +\sigma \int_{t}^{t+\tau} e^{-\theta (\tau-s)} dW_s \label{equ:ou_analytics}
\end{align}
Equation \ref{equ:ou_analytics} resembles the auto-regressive (AR) process form:
\begin{align}
	X_{t+\tau}=A+BX_t+C\varepsilon_{t+\tau}
\end{align}
with the constant part $A=(1-e^{-\theta \tau})\mu$, the slope part $B= e^{-\theta \tau}$, 
and the random part $ C =\sigma \sqrt{\frac{1-e^{-2\theta \tau}}{2\theta}}, 
\varepsilon \backsim N(0,1).$

In the simulation of instrumental effects, we propose a dual-pronged approach. 
The initial strategy involves utilizing the measured output with the 
local oscillator (LO) activated, yet in the absence 
of any imposed incoming signals, to establish a baseline representative of background noise. 
Subsequently, a predefined signal is superimposed upon this baseline, thereby 
emulating the response output during the execution of on-source observations. 
The alternative strategy employs the above-mentioned OU process model, 
which is tailored by assigning discrete parameter values  
$X_{i0}$, $\mu_i$, $\sigma_i$ and $\theta_i$. This customization can 
effectively simulate the variegated channel output of HSTDM. 

\subsection{ Space environment simulation}
\label{subsect:spaced_e}
In simulating the space environment, we primarily consider the following three aspects:
\begin{enumerate}[1)]
\item The coupling coefficient of the HSTDM's antenna with the same 
target source is different between the sunlit area and the shaded area;
\item Cosmic ray interference in the frequency range of 410-510 GHz;
\item Cosmic Microwave Background (CMB) noise; 
\end{enumerate}
To simulate the variable coupling coefficients of HSTDM's antenna in both sunlit 
and shaded regions, we employ a binary analytical approach. 
This methodology commences with the determination of the HSTDM's relative 
position with respects to the Sun and the Earth. Based on this 
positional assessment, the coupling coefficient is assigned a 
value of $\eta_1$ for periods when the antenna is exposed to sunlight and $\eta_2$
during periods of shadow. 
Consequently, this process generates a temporal sequence of coefficient values that correspond 
with the orbital data, as delineated in the parameters setting block of the simulation.

For the cosmic ray interference simulation, we assume that the cosmic ray interferences 
are transient in time and narrow banded, we therefore generate a narrowband spectral signal 
in the frequency range of 410-510 GHz with amplitude level and occurring timestamp  
adjustable through parameter setting specified in the parameter setting block of the simulation. 

For the CMB spectrum, we generate a white noise spectrum with an amplitude value of 2.7K.

\begin{figure*}[ht!]
	\begin{center}
		\begin{tabular}{c}
	\includegraphics[width=\textwidth]{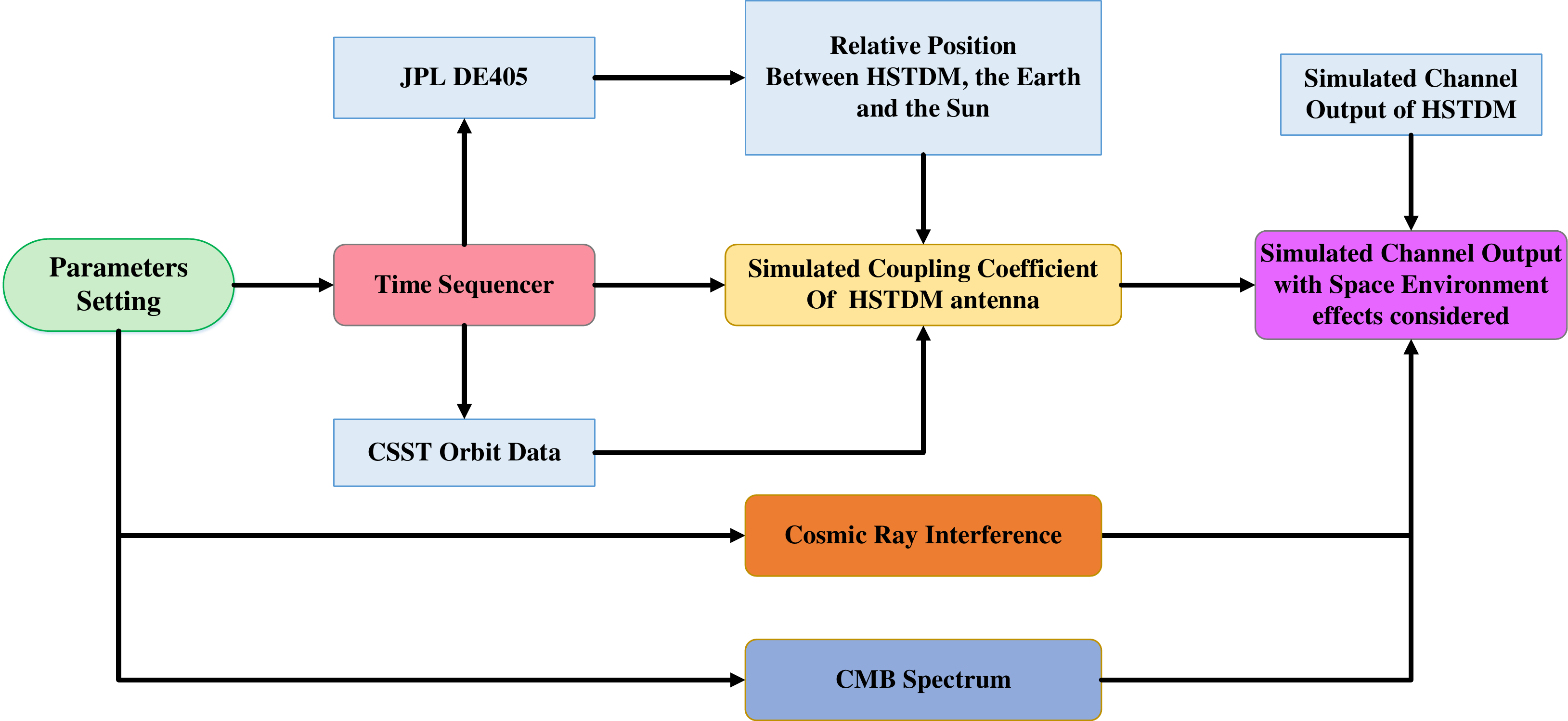}
\end{tabular}
\end{center}
	\caption{The schematic diagram of the space environment effects on HSTDM's output data.
	 In this diagram, three different space environment effects are 
	considered. The parameters defining and delineating these effects are encompassed within  
	the parameters setting block. The time sequencer is responsible for generating each time event that
	constitutes the observation process of HSTDM. The simulated channel output of HSTDM is 
	integrated with the simulated instrumental effects.
	}
    \label{fig:space_environ_simu}
\end{figure*}

The schematic diagram of the simulation of the space environment effects on the output 
data of HSTDM is illustrated in Fig \ref{fig:space_environ_simu}.
\subsection{Observation data flow simulation based on observing mode scenarios}
\label{subsect:observing_m}
The observing mode scenarios are essentially structured sequences of temporal events that 
constitute the framework of standard observational cycles. These events can be delineated 
into discrete operational segments, each can be described by a distinct set of parameters 
as elaborated in Table \ref{tab:single}, Table \ref{tab:OTF}, and equation \ref{equ:toff}. 

To accurately simulate the data flow generation for typical target mode or OTF observations, 
it is imperative to initially configure all pertinent parameters along the timeline, 
which define each operational step. These parameters encompass:
\begin{enumerate}[1)]
\item \textbf{Observation Parameters}: These specify the time slots allocated for all 
procedures conducted during a standard target mode or OTF observation. 
For a comprehensive understanding, refer to Table \ref{tab:single} and Table \ref{tab:OTF} 
for the definitions and meanings of these parameters.
\item \textbf{Time Sequencing Parameters}: These establish the start and end timestamps for the 
simulation, as well as the minimum duration of the discretized time slot, known as the time 
slot granularity.
\item \textbf{CSST Orbital Data}: Comprising millions of rows of positional and velocity data spanning a decade, 
with each row representing a 60-second interval. Interpolation is essential to derive positions 
and velocities at a time granularity finer than 60 seconds.
\item  \textbf{Space Environmental Parameters}: As previously discussed in subsection 
\ref{subsect:spaced_e}, these parameters play an important role in the simulation. 
\item  \textbf{Source Spectrum and Instrumental Effects Parameters}: These delineate the 
standard spectrum of the target source for target mode observation or spectral 
map of the target region for OTF observation, as well as the instrumental effects 
specific to the HSTDM. For target mode observations, standard spectrum data, 
accessible from existing archives, may require interpolation or extrapolation to 
align with the HSTDM's spectral data due to potential differences in the spectal 
channel numbers. 
In the context of OTF observation, it is imperative that the spectral data simulated 
at each timestamp undergoes initial interpolation from the standard spectral map data 
according to the most recently updated positional data. 
This preliminary step is essential prior to performing any subsequent interpolation 
or extrapolation to align with the HSTDM's ouput spectral data.   
The instrumental effects parameters are utilized as described in 
subsection \ref{subsect:instrument_e}.
\item  \textbf{Cold Load Data}: The chop load data as previously 
discussed in section \ref{sect:Obs}.  
\end{enumerate}

\begin{figure*}[ht!]
	\begin{center}
		\begin{tabular}{c}
	\includegraphics[width=\textwidth]{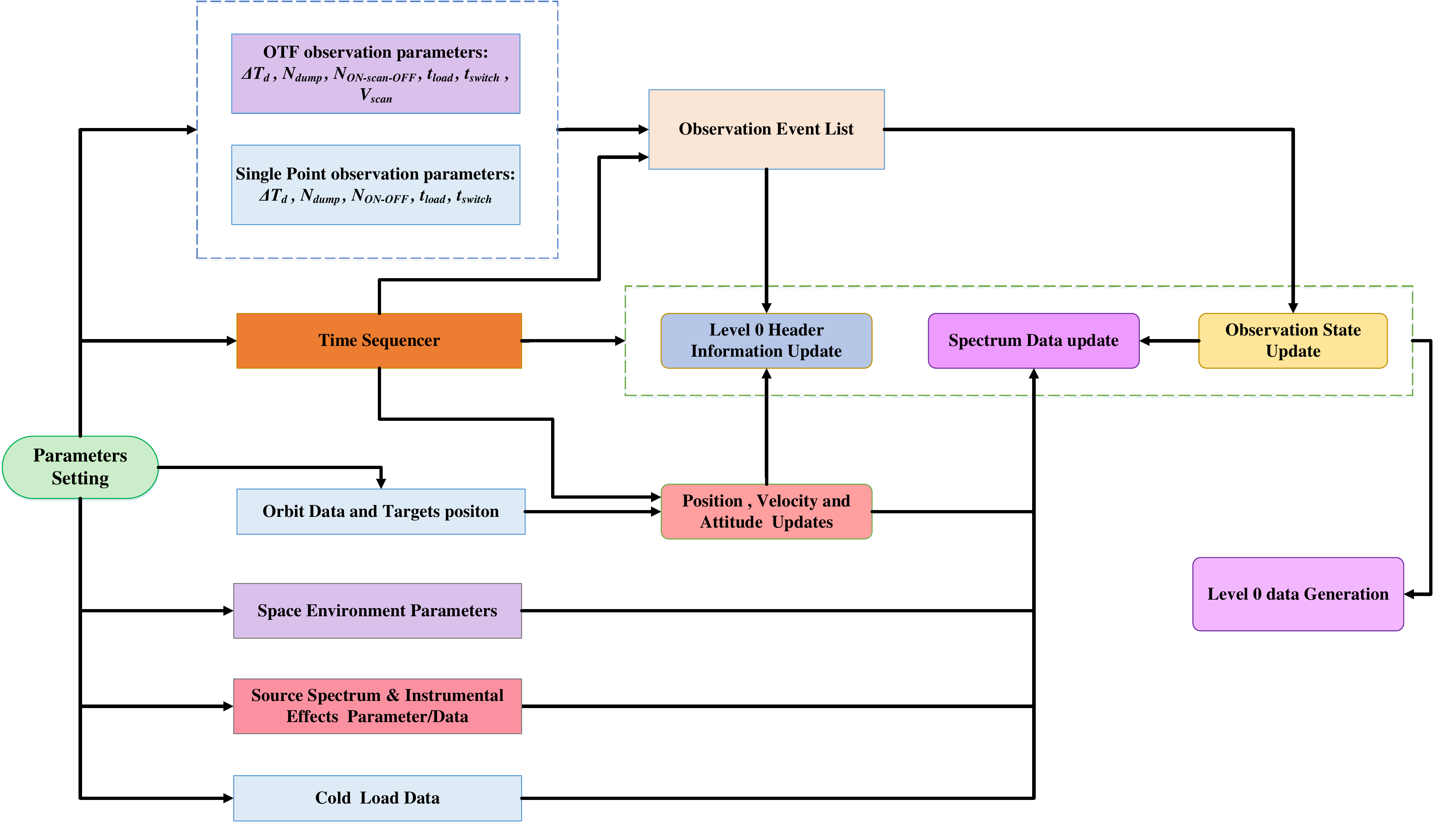}
\end{tabular}
\end{center}
	\caption{The schematic diagram of the data flow 
	simulation based on observing mode scenarios.
		In the presented diagram, two observational mode scenarios are taken into account, 
		alongside space environmental and instrumental effects.
	 The parameters setting block encompass parameters that 
	 define and distinguish specific observational modes, space environment and instrumental effects.
	 The time sequencer is responsible for generating each time event that
	constitutes the observation process of HSTDM under specific observing mode. 
	Additionally, level 0 header information is updated with each time event.
	}
    \label{fig:simu_data_flow}
\end{figure*}

Subsequently, employing the time sequencer and observation parameters, 
we generate a timeline of observation events at each time slot. 
We then iterate through the observation event list in each time slot, incrementally updating 
from fundamental information—such as CSST's position, velocity, attitude, and 
the observation state—to more sophisticated data, including space environmental impacts and 
observed frequency shifts in the spectral data. 
This process results in updated spectral data that incorporate all time-variant orbital and 
instrumental effects.

The spectral data, to which the aforementioned effects have been applied, 
in conjunction with the updated header information for the Level 0 
data and the updated observation state, 
collectively contribute to the generation of the standard Level 0 data.
This iterative process continues until the end of the observation event list.

We set the time granularity for observation events list be 0.25s, 
which is the minimum readout time of HSTDM during observations. 
The length of the observation event list depends on the time span of the simulated scenario.
It is important to note that the readout of HSTDM is in the form of spectral count while the 
scientifically usable data (source spectrum data) is in antenna temperature. 
Therefore, conversion from antenna temperature to spectral count is needed during 
the course of the spectrum data simulation. We choose  HSTDM's output data under 
hot and cold load operations, collected in the hardware's integrated test as our benchmark 
for converting antenna temperature to spectral count.
This conversion can be given by the following equation.
\begin{align}
	p_{sim} = \frac{T_{sim}-T_{c}}{T_{h}-T_{c}}(p_{h}-p_{c})+p_{c}
\end{align}
where $p_{sim}$ is the spectral data to be generated in spectral count, 
$T_{sim}$ is the antenna temperature of the simulated spectrum, 
$T_{c}$ and $T_{h}$ are the  black body temperature of cold and hot loads, respectively,
 $p_{c}$ and $p_{h}$ are the spectral counts measured during cold and hot load test of HSTDM.

The overview of the proposed data flow simulation is illustrated in Fig \ref{fig:simu_data_flow}

\section{ Simulation experiment and discussion}
\label{sect:simulation_discussion}
We implement the above mentioned simulation method in Python, and construct observation 
scenarios to generate observation simulation data accordingly.
The simulation data is a series of level 0 data files, which try to resemble the observational 
data of HSTDM as real as possible.

We feed those data into our developing HSTDM pipeline and get subsequent level 1 and level 2 data 
product. 
By comparing the spectrum part in the data product with that calculated in 
other verified approach, 
we can draw conclusions as to the accuracy of the HSTDM pipeline and the 
practicality of the proposed simulation method.

\begin{figure*}[t!]
	\begin{center}
		\begin{tabular}{c}
	\includegraphics[width=0.9\textwidth,trim=0cm 2cm 0cm 0cm]{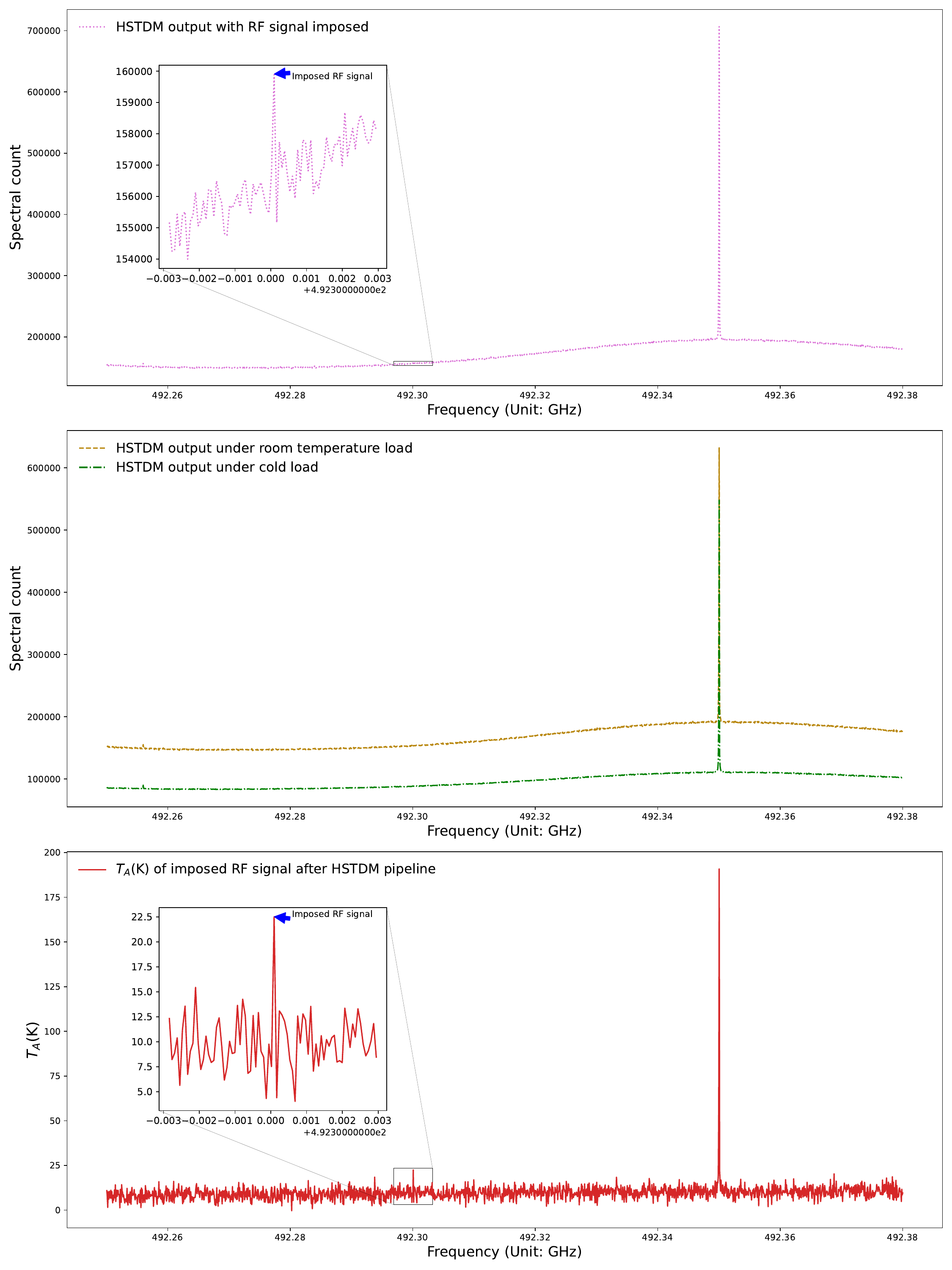}
\end{tabular}
\end{center}
	\caption{The simulation results of one target mode observation.
	The top subplot illustrates the channel output of HSTDM when 
	subjected to a faint Radio Frequency (RF) signal with a known frequency of 492.3GHz.
	The middle subplot presents the channel output of the HSTDM under both ambient room temperature (hot load) 
	and cryogenic (cold load) conditions.
	The bottom subplot depicts the RF signal recovered in antenna temperature units as a function of frequency, 
	as processed through the HSTDM Level 1 pipeline.
	The channel outputs are expressed in spectral count units, with the corresponding 
	channel numbers translated into observed frequencies displayed on the horizontal axes of 
	all three subplots.
	It is crucial to note that the horizontal axes of all three subplots are aligned to the 
	same observed frequency range, corresponding to channel numbers from 3414 to 5189.
	}
    \label{fig:single_plot}
\end{figure*}

We firstly build a target mode observation 
scenario based on the HSTDM output data collected during the brightness temperature 
tests in the integration testing phase of HSTDM qualification components. 

The collected data can be classified into 3 types:
\begin{enumerate}[1)]
\item \textbf{Cold Data:} data collected during cold load ($\approx $ 77K) operation;
\item \textbf{Room Temperature Data:} data collected during room temperature load 
($\approx $ 293.5K) operation;
\item \textbf{Signal Imposed Data:} data collected when a narrowband Radio 
Frequency (RF) signal is imposed on the antenna.   
\end{enumerate} 

By taking the cold data as the chop load data during nominal observation, the room 
temperature data as the output during the OFF position integration and the signal 
imposed data as the output during the source position
integration, we build a simple target mode observation scenario with observation data flow.
We then pass these data into the HSTDM pipeline as illustrated in Fig \ref{fig:hstdm-pipeline}
To better compare the spectrum data yielded by the HSTDM pipeline 
with that obtained through hot-cold calibration used in the brightness temperature tests\citep{JinjD_l},
 we turn both the frequency calibration (Doppler frequency) 
module and pointing calibration module off in the HSTDM Level 1 pipeline. 

Fig \ref{fig:single_plot} illustrates the spectral plots as 
follows: the top subfigure depicts the channel output of the HSTDM when a faint radio frequency is 
applied ; the middle subfigure presents the channel output of the HSTDM under both hot and cold 
load conditions; and the bottom subfigure displays the spectrum data derived from the HSTDM Level 1 
 pipeline. Given that the imposed RF signal is a faint single-frequency RF with a known frequency 
 of 492.3GHz,
 its corresponding channel output is barely discernible over a broad channel range.
  We opt for zoomed plot around its corresponding channel number, as shown in the zoom region 
  of the top subfigure.
 Additionally, the recovered RF signal, along with its measured antenna temperature, is 
 displayed in the zoom region of the bottom subfigure.

It is obvious from the figure that we can obtain the antenna temperature $T_A$(K) of 
the imposed RF signal after the routine process of HSTDM pipeline, though we turn the 
pointing calibration module and frequency calibration module off. 
The spectral spike, as clearly displayed through all the subplots 
in Fig \ref{fig:single_plot} is perhaps caused by the IF chain noises.
The intensity of the spectral spike is much larger than that of imposed RF signal, 
as observed in the top and bottom subfigures.

We measure the antenna temperature $T_A$ of the RF signal be 22.53104 K.
For comparison, $T_A$ calculated using an alternative method for the brightness 
temperature test also yields 22.53039 K. This alternative approach, as elaborated 
in reference \citet{JinjD_l}, involves leveraging the approximate linear relationship between  
antenna temperature of the imposed signal source and the channel output.
This linear relationship is mathematically articulated in equation \ref{equ:hot-cold}. 
This alternative approach, leveraging measurement data obtained from both cold 
and hot load conditions to conduct a fitting procedure, is fundamentally analogous 
to the core module within the HSTDM Level 1 pipeline, albeit distinct in its 
approach to processing the measurement data. 
Both values are in good agreement, albeit not exactly the same.
The minor differences in the calculation result stems from the 
specific details in how the two methods process the measurement data.
Therefore, we have roughly verified the accuracy of the HSTDM level 1  pipeline. 

\begin{figure*}[tbh!]
	\begin{center}
		\begin{tabular}{c}
	\includegraphics[width=0.9\textwidth]{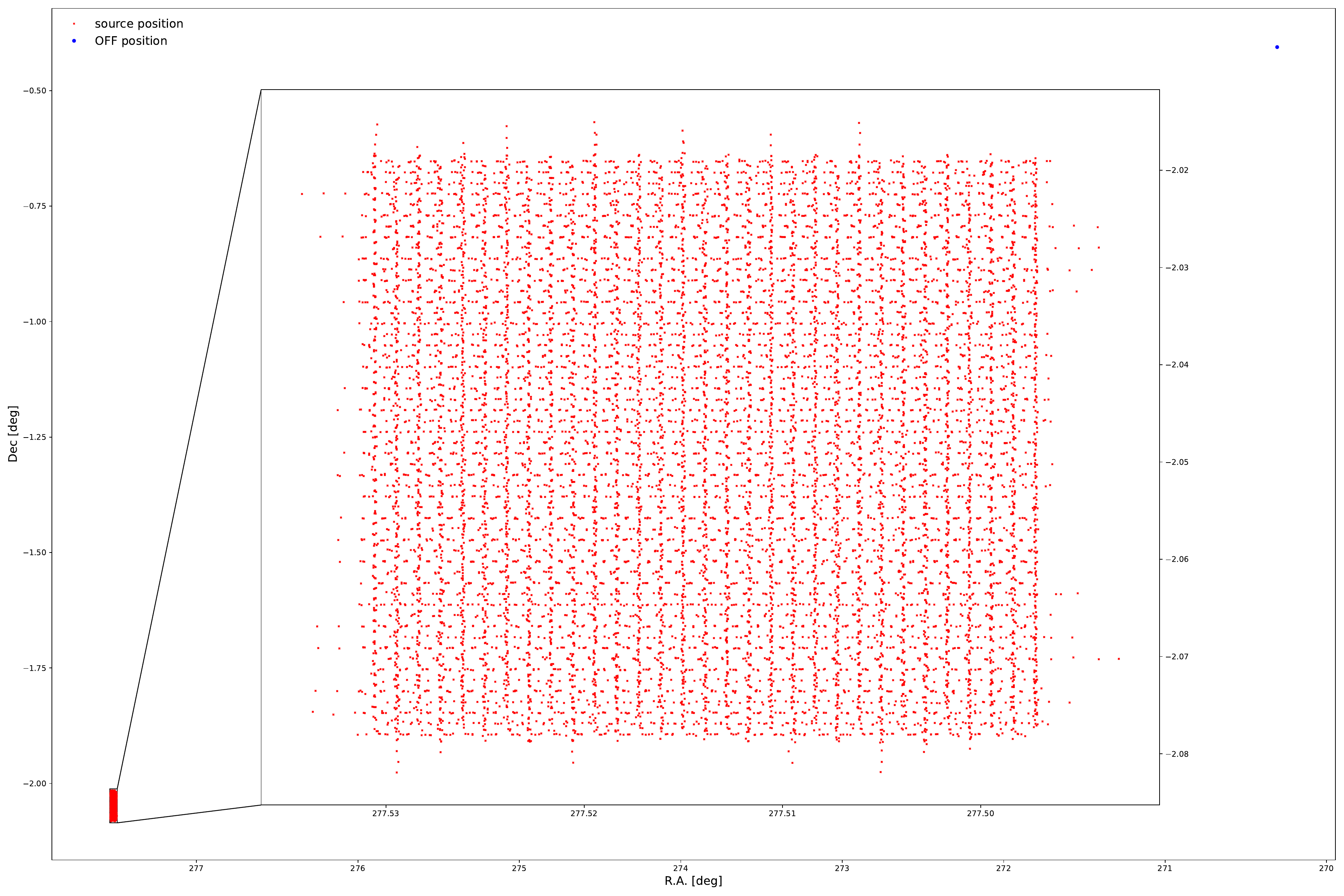}
\end{tabular}
\end{center}
	\caption{The distribution of the observation point of the simulated OTF scenario.
	Source map is filled with Source position points, which are indicated by red crosses in the 
	zoom region of the plot. In contrast, the OFF position point is denoted by a  
	blue circle. Collectively, there are a total of 26,478 source position points  
	in the simulated scenario.}
    \label{Fig: OTF_point}
\end{figure*}

\begin{figure*}[tbh!]
	\begin{center}
		\begin{tabular}{c}
	\includegraphics[width=\textwidth]{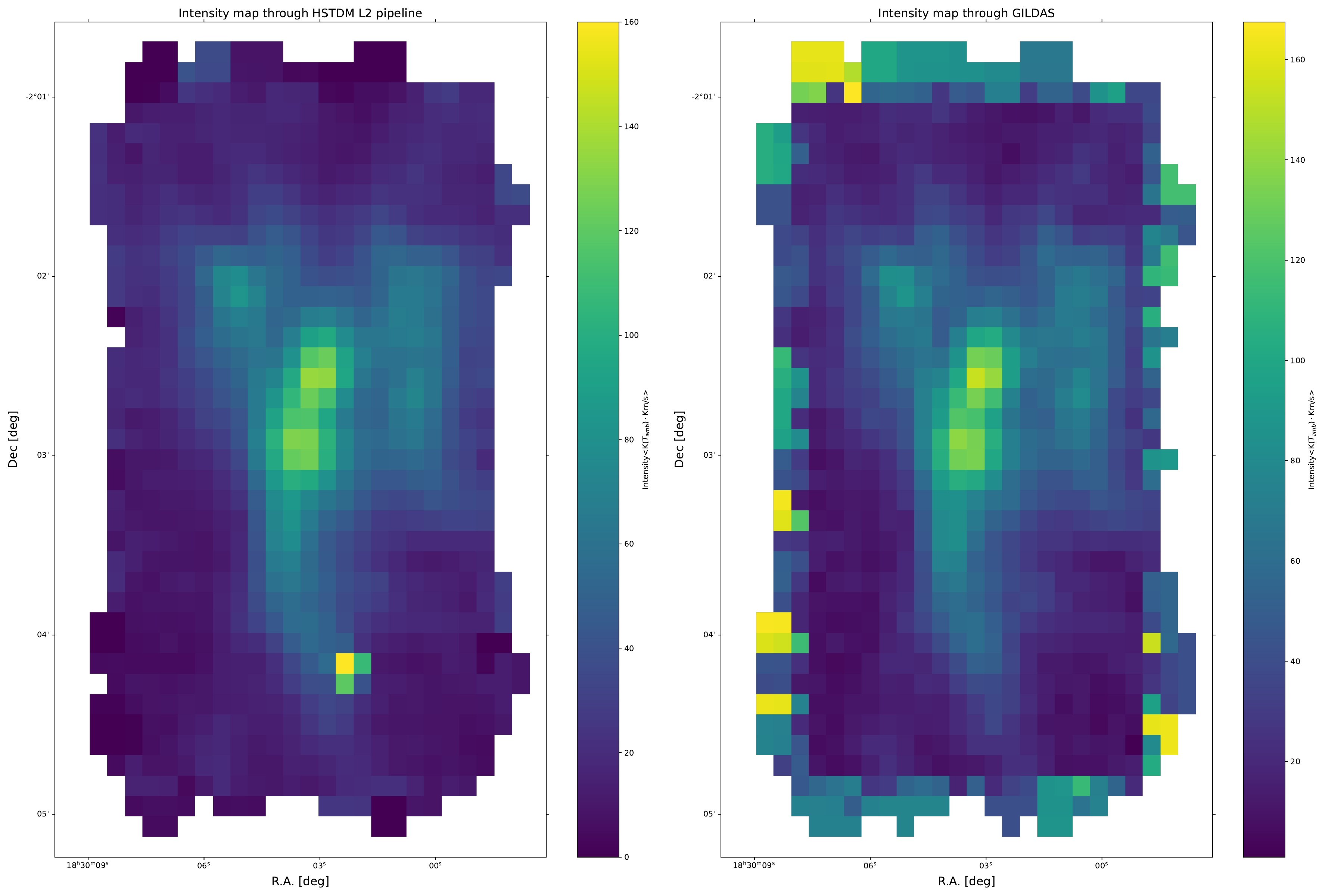}
\end{tabular}
\end{center}
	\caption{The Comparison of intensity map of CO(4-3) 
	derived through two distinct methods.
	 The left plot represents intensity map generated by using 
	 Regridding module within HSTDM Level 2 pipeline.
	 In contrast, the right plot shows the intensity map obtained through 
	 the utilization of the GILDAS software.}
    \label{Fig: OTF_results_compare}
\end{figure*}

We also generate a series of simulation data files based on an 
archived OTF observation scenario 
by focusing on three critical aspects. 
Initially, we obtain the APEX spectral map data of the CO(4-3) emission line 
for SERP-MM18 as part of an effort to further investigate the outflows from Serpens 
South, complementing previous studies on its chemistry \citep{2023AA679A39G}. 
This dataset provides both the off-position spectral data
 and on-source spectral data after subtracting the off position spectral data.
 Subsequently, we convert the header information and the data dimension of each 
 data file into the HSTDM Level 1 data format. 
 Ultimately, we utilize these data files as inputs for the Regridding module within the 
 HSTDM Level 2 pipeline, which produces the integrated intensity map of the CO (4-3) 
 emission line. We then compare this intensity map with one generated by 
 a verified pipeline using the GILDAS software\citep{glidas_web}.

It is important to note that the generation of the aforementioned OTF simulation data 
encompasses solely the 
conversion of  FITS header information and data dimension.
We opt for this approach primarily because our objective is to validate the 
 Regridding module within the HSTDM Level 2 pipeline.
 
Fig \ref{Fig: OTF_point} depicts the distribution of observation points within the simulated OTF 
scenario.The source map is populated with source position markers, which are distinctly indicated 
by red crosses in the zoom section of the plot. In stark contrast, the OFF position is denoted 
by a blue circle. Collectively, there are a total of 26,478 source position points accounted 
for in the simulation.

Fig \ref{Fig: OTF_results_compare} displays two distinct subplots: the left subplot 
illustrates the intensity map of CO(4-3) derived through the Regridding module within
the HSTDM Level 2 pipeline. In contrast, the right subplot represents intensity map 
generated by using verified GILDAS pipeline. Both methods use the
same sampled data files, though different only in header format and data dimension of each
data file.
Upon comparing the intensity maps generated by these two methods, it is observed that 
the intensity values are largely consistent across corresponding positions within the scanned 
rectangular region. Additionally, discrepancies in intensity are noted along the border 
regions of the scanned area.
Upon examining the reasons behind these discrepancies, we surmise that the variances 
in the regridding algorithms, encompassing the respective parameters of both methods, 
could potentially account for the observed differences.

While we have thus far implemented only the essential part 
of the proposed simulation method to generate Level 0 data for the HSTDM Level 1 pipeline,
and Level 1 data for the Regridding module within the HSTDM Level 2 pipeline, 
we have already carried out preliminary validations to ascertain the accuracy 
of the HSTDM Level 1 pipeline, as well as the Regridding module 
within the HSTDM's Level 2 pipeline.
This initial validation underscores the practical utility of our proposed simulation approach. 
Moving forward, we are committed to developing the remaining aspects of the proposed 
simulation method and will persistently enhance them through integration with the 
most recent ground-based measurement data 
and the evolving testing requirements of the HSTDM pipeline.

\section{Conclusions}
\label{sect:conclusion}
In this paper, we investigate synthetic data generation methods for HSTDM observations.
 Firstly, we parameterize 
the basic operations in the observational cycles for both the target mode observation and 
the OTF observation. Secondly, we propose a simulation framework that focuses on 
emulating the instrumental effects and space environment, as well as generating data 
streams based on observational modes. 
We provide detailed introductions to these simulation methods which altogether strives to mimic 
the actual output of HSTDM as closely as possible. Thirdly, we implement 
the simulation method and conduct simulation experiment, thereby validating  
practical applicability  of the proposed simulation method.

\section{Acknowledgment}
The authors wish to express their gratitude to Dr. YouHua Xu from the National Astronomical Observatory,
 Chinese Academy of Sciences, for generously supplying the orbital data of the CSST platform,
  which has been pivotal to our research endeavors.
Sincere appreciation is extended to the HSTDM hardware development team for their invaluable contribution 
of test data during the system integration testing phase. Their dataset has undeniably served as the 
foundational bedrock upon which the research presented in this article is built.
We also extend our heartfelt thanks to the Visiting Scholar Gong Yan from Purple Mountain Observatory, 
Chinese Academy of Sciences, for providing the CO(4-3) spectral data, 
which are instrumental in conducting the OTF simulation analysis presented in this article.  
Additionally, we acknowledge the guidance and leadership of the CSST science data processing group, 
as well as the insightful discussions with many distinguished members of this group, which have greatly 
enhanced the depth and quality of our work.
This article is made possible under the support of the Natural Science Foundation of China (NSFC) under 
the grants 12427901 and 12403095, and the support of the program of the 
Ministry of Science and Technology of the People's Republic of China under the grant 2023YFA1608200.
It is important to mention that some results in this paper have been derived using the Astropy, 
NumPy, and pandas packages under the Python programming environment, we thank the developers of the 
Python programming language and the maintainers of these packages for making their code available on a free 
and open-source basis.
\section{Data availability}
The data underlying this paper will be shared on reasonable request to the corresponding author.

\bibliographystyle{raa}
\bibliography{sytan_bib} 

\label{lastpage}

\end{document}